\documentclass[longnamesfirst,11pt,preprint]{aastex}
\usepackage{psfig}
\received{2000 November~16}
\revised{2001 May~14}
\accepted{2001 May~23}
\paperid{MS~53027}
\shorttitle{W49B Free-free Absorption}
\shortauthors{Lacey et al.}
\newcommand{\kms}{\mbox{km~s${}^{-1}$}}
\newcommand{\mjybm}{\mbox{mJy~beam${}^{-1}$}}
\newcommand{\cd}{\mbox{cm${}^{-2}$}}

\begin{document}

\title{Spatially-resolved Thermal Continuum Absorption against the
	Supernova Remnant W49B}

\author{C.~K.~Lacey\altaffilmark{1}}
\affil{National Research Council and Naval Research Laboratory,
	Code~7213, Washington, DC  20375-5351 USA}
\email{lacey@rsd.nrl.navy.mil}
\altaffiltext{1}{Present address: Department of Physics and Astronomy,
	University of South Carolina, Columbia, SC  29208 USA}

\author{T.~Joseph~W.~Lazio, Namir~E.~Kassim} 
\affil{Naval Research Laboratory, Code~7213, Washington, DC 
        20375-5351 USA}
\email{lazio@rsd.nrl.navy.mil}
\email{kassim@rsd.nrl.navy.mil}

\author{N.~Duric}
\affil{Department of Physics and Astronomy, University of New Mexico,
	800 Yale Boulevard NE, Albuquerque, NM  87131 USA}

\author{D.~S.~Briggs\altaffilmark{2}}
\affil{National Research Council and Naval Research Laboratory,
	Code~7210, Washington, DC 20375-5351 USA}
\altaffiltext{2}{Deceased}

\and

\author{K.~K.~Dyer}
\affil{North Carolina State University, Raleigh, NC 27695 USA}
\email{Kristy$\_$Dyer@ncsu.edu}

\begin{abstract}
We present sub-arcminute resolution imaging of the Galactic supernova
remnant \objectname[]{W49B} at~74~MHz (25\arcsec) and~327~MHz
(6\arcsec), the former being the lowest frequency at which the source
has been resolved. While the 327~MHz image shows a shell-like
morphology similar to that seen at higher frequencies, the 74~MHz
image is considerably different, with the southwest region of the
remnant almost completely attenuated. The implied 74~MHz optical depth
($\approx 1.6$) is much higher than the intrinsic absorption levels
seen inside two other relatively young remnants, Cas~A and the Crab
Nebula, nor are natural variations in the relativistic electron energy
spectra expected at such levels. The geometry of the absorption is
also inconsistent with intrinsic absorption.  We attribute the
absorption to extrinsic free-free absorption by a intervening cloud of
thermal electrons. Its presence has already been inferred from the
low-frequency turnover in the integrated continuum spectrum and from
the detection of radio recombination lines toward the remnant. Our
observations confirm the basic conclusions of those measurements, and
our observations have resolved the absorber into a complex of
classical \ion{H}{2} regions surrounded either partially or fully by
low-density \ion{H}{2} gas.  We identify this low-density gas as an
extended \ion{H}{2} region envelope (EHE), whose statistical
properties were inferred from low resolution meter- and
centimeter-wavelength recombination line observations. Comparison of
our radio images with \ion{H}{1} and H${}_2$CO observations show that
the intervening thermal gas is likely associated with neutral and
molecular material as well.  This EHE may be responsible for the
enhanced radio-wave scattering seen in the general direction of the
\objectname[]{W49} complex.
\end{abstract}

\keywords{ISM: individual (W49B) --- ISM: structure --- radio lines:
	ISM --- scattering --- supernova remnants}

\section{Introduction}\label{sec:intro}

The integrated radio continuum spectra of Galactic supernova remnants
(SNRs) are generally power-law from meter to centimeter wavelengths
and shorter, with a continuum of spectral indices ranging from $\alpha
\simeq -0.7$ ($S \propto \nu^\alpha$) for shell-type remnants to
$\alpha \simeq -0.1$ for plerions, with many presenting blended
emission at intermediate indices.  However, below~100~MHz, roughly
two-thirds of SNRs show spectral turnovers indicative of thermal
absorption \citep[][and references within]{ds75,k89}.  The inferred
continuum optical depths and the poor correlation between the presence
of a turnover and the, albeit poorly constrained, distance to a SNR are
inconsistent with absorption arising in the (globally-distributed)
warm ionized medium \citep[WIM, $n \sim 0.1$~cm${}^{-3}$,][]{kh88}.  Instead, the
absorption must arise from localized ionized regions having an
enhanced density ($n \gtrsim 1$~cm${}^{-3}$) but of unknown scale size
and with a small ($\le 1$\%) filling factor \citep{k89}. The favored
interpretation has been that the absorbers represent extended
\ion{H}{2} region envelopes (EHEs), ionized gas surrounding normal
\ion{H}{2} regions, as postulated by \cite{a86} from comparison of
centimeter- and meter-wavelength radio recombination line (RRL)
observations.  Alternatively, the absorption could be caused by the
superposition of many small, normal \ion{H}{2} regions or planetary
nebulae along the line of site.  Until now, low-frequency observations
have had insufficient resolution to resolve spatially the morphology
of the absorption and thereby constrain the geometry and scale size of
the absorber.

We distinguish between the free-free absorption responsible for
low-frequency turnovers and the \emph{intrinsic} thermal absorption
now documented in two Galactic SNRs (\objectname[]{Cas~A},
\citealt{kpde95}; \objectname[]{Crab Nebula}, \citealt{bkfpeh97}).  In
these two SNRs, the free-free optical depths in their interiors are
far smaller than the levels considered here, and the radio absorption
can be linked to thermal material seen at other wavelengths, e.g.,
optical filaments in the Crab Nebula.

The supernova remnant \objectname[]{W49B} (\objectname[]{G43.3$-$0.2})
is the nonthermal component of the \objectname[]{W49} complex and is
approximately 12\arcmin\ from the \ion{H}{2} region
\objectname[]{W49A}.  It is a relatively young ($\sim 3000$~yr,
\citealt{ptbs84,spjp85}), bright ($\approx 40$~Jy at~1~GHz), and
reasonably extended ($\approx 5\arcmin$) shell-type \hbox{SNR}.  As
such it has been a common target for testing the predictions of shock
acceleration theory against the energy spectrum of the relativistic
electrons inferred from spectral index studies above~100~MHz
\citep{mr94,dkrp00}.  Its integrated spectrum also shows a
low-frequency turnover near~100~MHz \citep{ds75,k89}.  \cite{k89}
fitted its spectrum with a power law, finding a spectral index~$\alpha
= -0.4$ above~100~MHz and a free-free 30.9~MHz optical depth of
$\tau_{30.9} = 0.9 \pm 0.3$.  Low resolution detection of RRLs in the
direction of \objectname[]{W49B}, unusual for a Galactic SNR, provide
further strong evidence for ionized gas along the line of sight
\citep{pd76}. Its distance is estimated to be 10~kpc based on
\ion{H}{1} absorption \citep{rgmb72}. Though there is some dispute
about whether these two sources are physically related, there is
general agreement that \objectname[]{W49A} is \emph{behind}
\objectname[]{W49B} \citep[e.g.,][ but see \citealt{bt00}]{rgmb72}.

This paper reports the first low frequency radio observations to
resolve spatially the long-inferred, extrinsic thermal absorption
toward a Galactic \hbox{SNR}. We describe the observations in
\S\ref{sec:observe} and present the results in \S\ref{sec:results}. We
discuss the implications of our observations with respect to the
properties of the absorber in \S\ref{sec:discussion} and summarize our
conclusions in \S\ref{sec:conclude}.

\section{Observations and Data Reduction}\label{sec:observe}

We used the VLA to observe \objectname[]{W49B} at~74 and~327~MHz.
Table~\ref{tab:observelog} summarizes various observing details.

\begin{deluxetable}{lcccccc}
\tablecaption{VLA Observing Log\label{tab:observelog}}
\tabletypesize{\footnotesize}
\tablewidth{0pc}
\tablehead{
	& \colhead{VLA} \\
 \colhead{$\nu$} & \colhead{Configuration} & \colhead{Epoch} & \colhead{$\Delta\nu$}   & \colhead{$T$} 
	& \colhead{Beam} & \colhead{$\Delta I$} \\
 \colhead{(MHz)} & & & \colhead{(MHz)} & \colhead{(hr)} 
 	& \colhead{(\arcsec\ $\times$ \arcsec)} & \colhead{(\mjybm)}
}
\startdata

\phn74 & A & 1998 February~22 & 1.4 & 5.0 & $26 \times 23$ @ $-10\arcdeg$ & 90   \\
327    & A & 1995 August~21   & 1.6 & 5.9 & $6.6 \times 6.2$ @ 71\arcdeg  & 0.62 \\
327    & B & 1995 October~8   & 1.6 & 5.9 & \ldots             & \ldots \\
327    & B & 1994 August~7    & 3.1 & 3.4 & \ldots             & \ldots \\
327    & C & 1994 October~24  & 3.1 & 2.1 & \ldots             & \ldots \\

\enddata
\tablecomments{The 327~MHz observations were combined to produce a
single image.  We list the beam and rms noise level on that image only once.}
\end{deluxetable}

The 74~MHz observations were conducted on 1998 February~22 with the
VLA in the A configuration. We note some of the salient features of
the NRL-NRAO 74~MHz system on the \hbox{VLA}. Signals are received by
a pair of crossed dipoles mounted at the prime focus of the VLA's 25~m
diameter antennas. The signals are filtered to a 1.5~MHz bandwidth,
converted to right- and left-circular polarization, and amplified. As
the system temperature ($\sim 10^3$~K) is dominated by the Galactic
background, the receivers are uncooled. The filtered and amplified
signals are then passed to the regular VLA IF chain.  The VLA IF chain
can handle four simultaneous IFs. Two of these are utilized by the
right- and left-circularly polarized signals at~74~MHz. The system was
designed so that the remaining two IFs may also carry signals
at~327~MHz. In this case, the simultaneous 3~MHz bandwidth 327~MHz
data were acquired in the event that dual-frequency ionospheric phase
referencing \citep[DFIPR, ][]{kped93} was later required to compensate
for strong ionospheric effects. Often, as was the case here, DFIPR was
not required.

The 327~MHz observations occurred between 1994 and~1995 in the VLA's
A-, B-, and C-configurations (Table~\ref{tab:observelog}).  These
multi-configuration observations provided a range of spatial
frequencies and ensured that larger angular scale structures in the
SNR were well represented at the higher frequency.  In combining the
observations from the various configurations, we made no attempt to
compensate for any time-dependent changes in the \hbox{SNR}.

Post-processing of low-frequency VLA data uses procedures similar to those
at higher frequencies, though certain details differ.\footnote{%
A full description of low-frequency VLA data reduction procedures is at\hfil\\
\url{http://rsd-www.nrl.navy.mil/7213/lazio/tutorial/}.%
}  \objectname[]{Cygnus~A} served as the bandpass, flux density, and
initial phase calibrator. Several iterations of self-calibration were
then utilized to make the final images.

Two significant differences for the post-processing are the need to
remove radio frequency interference (RFI) and the wide-field imaging
requirement imposed by the large fields of view. In order to combat
RFI, the data were acquired with a much higher spectral resolution
than used for imaging. Excision of potential RFI is performed on a
per-baseline basis for each 10~s visibility spectrum. The large fields
(11\arcdeg\ at~74~MHz and~2\fdg5 at~327~MHz) and non-coplanar
baselines of the VLA require full three-dimensional deconvolution to
reach thermal limits. The 74 MHz data was reduced in AIPS using
\texttt{IMAGR}, while NRAO's Software Development Environment
polyhedron algorithm \texttt{dragon} \citep{cp92} was utilized
at~327~MHz.  The rms noise level in the resulting 327~MHz image,
0.62~\mjybm, makes this one of the most sensitive, low-frequency
images ever obtained by the VLA toward the inner Galaxy.

\begin{figure}[tbh]
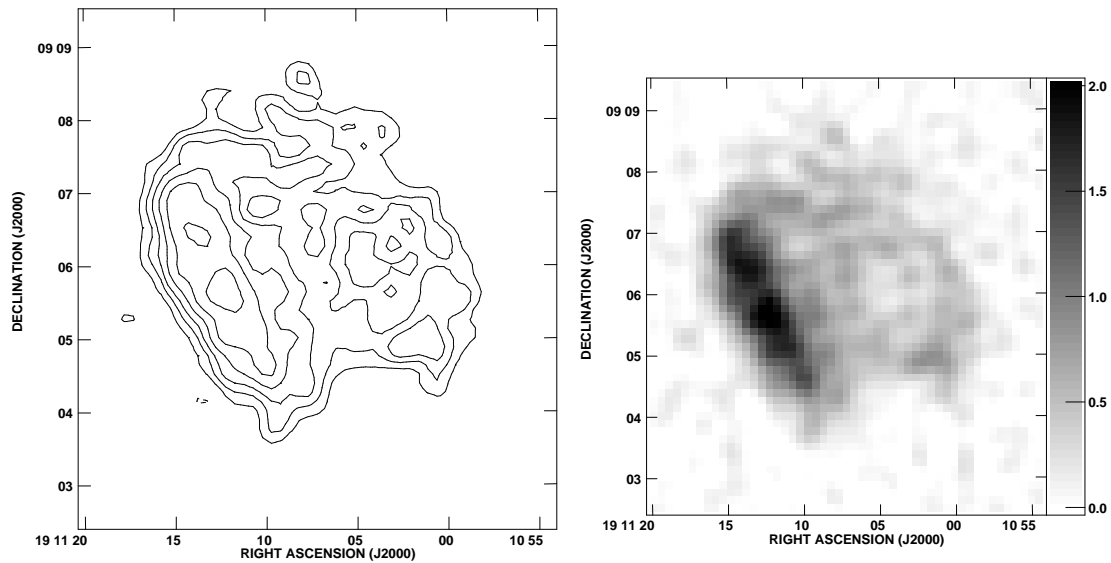

\begin{center}
 \mbox{\psfig{file=Lazio_f1a.ps,width=0.45\textwidth,silent=}
       \psfig{file=Lazio_f1b.ps,width=0.45\textwidth,silent=}}
\end{center}
\vspace{-1.25cm}
\caption[]{The supernova remnant \protect\objectname[W]{W49B}
at~74~MHz. 
(\textit{a})~The rms noise level in the image is
90~\mbox{mJy~beam${}^{-1}$}, and the contour levels are
90~\mbox{mJy~beam${}^{-1}$} $\times -3$, 3, 5, 7.07, 10, 14.1, 20. The
beam is $26\arcsec \times 23\arcsec$ at a position angle
of~$-10\arcdeg$.
(\textit{b})~The gray scale flux range is linear between~0
and~2~Jy~beam${}^{-1}$.}
\label{fig:74}
\end{figure}

\begin{figure}[tbh]
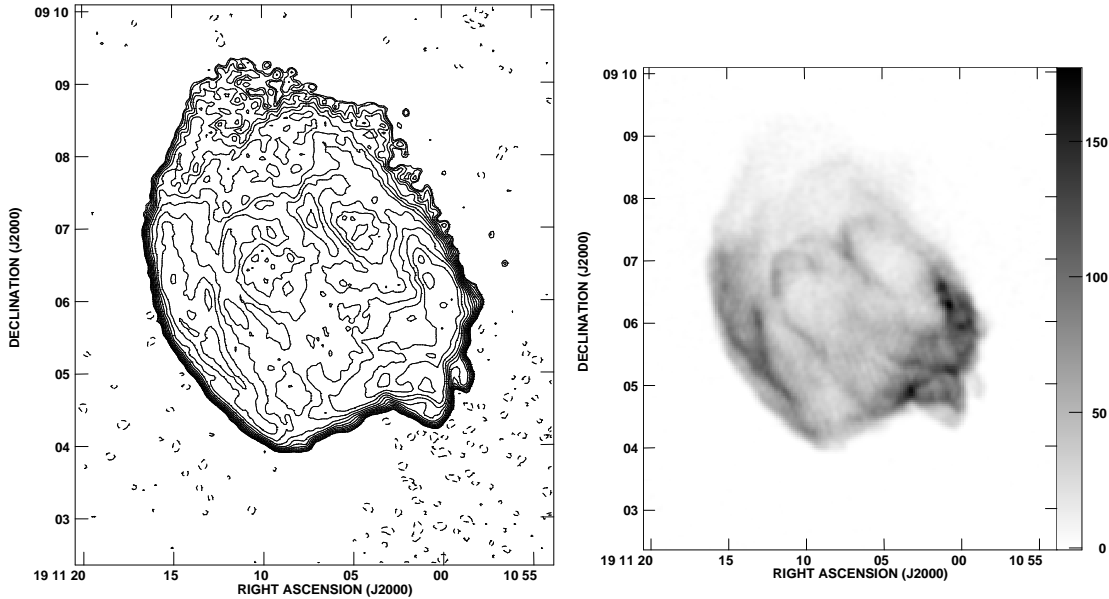

\begin{center}
 \mbox{\psfig{file=Lazio_f2a.ps,width=0.45\textwidth,silent=}
       \psfig{file=Lazio_f2b.ps,width=0.45\textwidth,silent=}}
\end{center}
\vspace{-1cm}
\caption[]{The supernova remnant \protect\objectname[W]{W49B}
at~327~MHz.  The sub-milliJansky sensitivity of this image makes it
one of the highest dynamic range, low-frequency images ever obtained
by the \hbox{VLA}.
(\textit{a})~The rms noise level in the image is
0.62~\mbox{mJy~beam${}^{-1}$}, and the contour levels are
$0.62~\mbox{mJy~beam${}^{-1}$} \times -3$, 4, 5, 7.07, 10, 14.1, 20,
\ldots. The beam is $6\farcs6 \times 6\farcs2$ at a position angle
of~71\arcdeg.
(\textit{b})~The gray scale flux range is linear between~0
and~175~Jy~beam${}^{-1}$.}
\label{fig:327}
\end{figure}

Figures~\ref{fig:74} and~\ref{fig:327} show the 74 and~327~MHz images
of \objectname[]{W49B}, respectively. The images are strikingly
different. At~327~MHz (Figure~\ref{fig:327}) \objectname[]{W49B} shows
a limb-brightened, shell morphology similar to that seen at higher
frequencies and in previous 90~cm observations \citep{mr94}, with the
southwest portion of the supernova remnant being the brightest.  In
contrast, at~74~MHz (Figure~\ref{fig:74}) the morphology is reversed,
with the eastern portion brightest, and the emission toward the
southwest dramatically suppressed.

\section{Results}\label{sec:results}

Figure~\ref{fig:74} suggests strongly that we have resolved spatially
the thermal absorption inferred previously from the low-frequency
turnover in the continuum spectrum and from RRLs detected toward
\objectname[]{W49B}.  In this section, we first demonstrate that the
74~MHz flux density we measure is consistent with the degree of
absorption inferred from the integrated continuum spectrum.  We then
argue that this level of absorption and its morphology favor an
extrinsic ISM absorber. Finally, we address the scale size constrained
by the spatially-resolved structure in the absorber toward
\objectname[]{W49B}.

\subsection{Integrated Spectrum and Absorption of W49B}\label{sec:spectrum}

\begin{figure}[tbh]
\begin{center}
 \mbox{\psfig{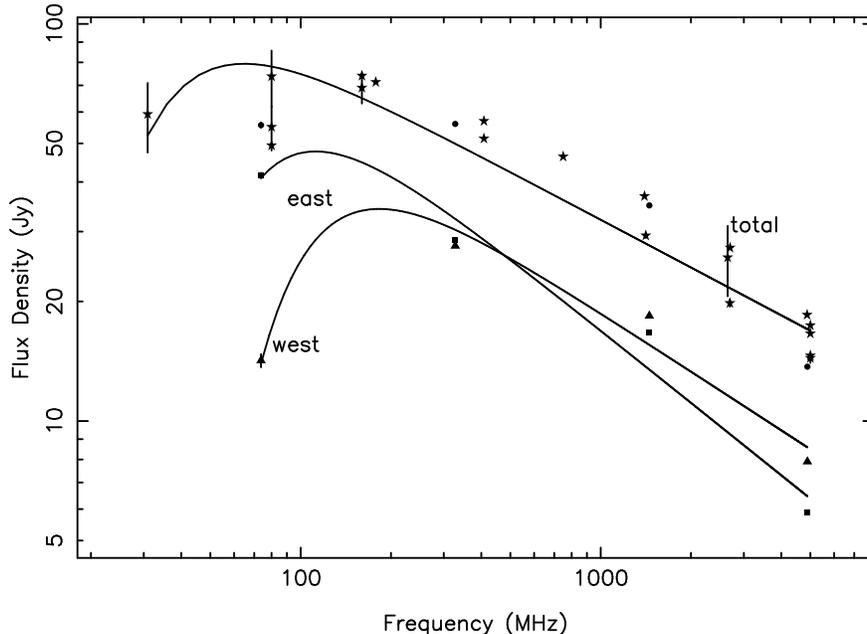}}
\end{center}
\vspace{-1cm}
\caption[]{The low-frequency spectrum of the supernova
remnant~\protect\objectname[W]{W49B}. The data for the total spectrum
are taken from \cite{k89} and are shown as stars.  The circles show
the total spectrum taken from the same images used to construct the
spectra for the eastern and western halves of the remnant.  The data
for the eastern (squares) and western (triangles) halves of the
remnant are taken from the images presented here and from the images
(at~1452 and~4885~MHz) presented by \cite{mr94}, with the demarcation
between the eastern and western halves described in the text
(\S\ref{sec:spectrum}).  The flux densities of the eastern and western
halves are nearly identical at 327~MHz.  Uncertainties on the flux
density measurements are shown, but in many cases the sizes of the
uncertainties are comparable to or smaller than the sizes of the
symbols.  The curve describing the total spectrum is taken from
\cite{k89}; those describing the eastern and western halves of the
remnant are fits resulting from the data shown, as discussed in the
text.}
\label{fig:spectrum}
\end{figure}

Figure~\ref{fig:spectrum} shows the spectrum of the total remnant and
compares our flux density measurements to those published previously.
The data for the spectrum of the remnant are taken from \cite{k89}, as
is the fit to those data.  The expected 74~MHz flux density from
\citeauthor{k89}'s~(\citeyear{k89}) spectral fit (66~Jy) exceeds the value we measure (55~Jy) by 20\%.  We do not consider this discrepancy to be significant.
First, the lowest frequency flux densities compiled by \cite{k89} were
often plagued by confusion, resulting from the poor angular
resolution, and therefore tend to overestimate the flux densities of
SNRs.  Variations at the levels of the uncertainties in the spectral
fits ($\sim -0.1$ in spectral index and $\approx 30$\% in optical
depth) can account for the discrepancy between expected and observed
flux densities.  Second, some of the discrepancy may result from
\objectname[]{W49B} being partially resolved out by our
A-configuration 74~MHz observations.  Even so, this would not affect
our general conclusions: There is considerable asymmetry in the
structure of \objectname[]{W49B} at~74~MHz, an asymmetry that differs
substantially from what is seen at higher frequencies.  Moreover,
B-configuration observations obtained recently show a similar
east-west asymmetry of the 74~MHz emission, albeit at lower resolution
(Perley~2000, private communication), giving us further confidence
that resolution effects are not the dominant cause of the
low-frequency asymmetry. We therefore consider our 74 MHz integrated
flux density consistent with previous work, although the optical
depths we present here could be considered conservatively as upper
limits.  Our multi-configuration observations at~327~MHz means that
resolution effects do not affect these data.

Figure~\ref{fig:spectrum} also shows the spectra for the eastern and
western halves of \objectname[]{W49B}. The data for the eastern and
western halves of the remnant were obtained from the images presented
here and from images presented by \citet[][at~1452 and~4885~MHz]{mr94}
and are summarized in Table~\ref{tab:spectrum}. We used the 74~MHz
image as a guide for delimiting the eastern and western halves of the
remnant, with the dividing line being the right ascension $\alpha =
19^{\mathrm{h}}11^{\mathrm{m}}06\fs6$. The flux densities shown for
the eastern and western halves of the remnant were taken to be the
integrated brightness of the remnant either east or west of this
dividing line. No attempt has been made to account for
frequency-dependent position shifts within the remnant.

\begin{deluxetable}{lccc}
\tablecaption{Spectrum of W49B\label{tab:spectrum}}
\tablewidth{0pc}
\tablehead{
 \colhead{$\nu$} & \colhead{Total}   & \colhead{East Half}  &
	\colhead{West Half} \\
 \colhead{(MHz)} & \colhead{(Jy)} & \colhead{(Jy)} & \colhead{(Jy)}
}
\startdata

\phn\phn74 &  55.6 $\pm$ 0.8  & 41.5 $\pm$ 0.7  & 14.2 $\pm$ 0.5 \\
\phn326    &  56.0 $\pm$ 0.02 & 28.5 $\pm$ 0.02 & 27.6 $\pm$ 0.02 \\
1452       &  34.9 $\pm$ 0.07 & 16.7 $\pm$ 0.05 & 18.4 $\pm$ 0.04 \\
4885       &  13.7 $\pm$ 0.03 &  5.87 $\pm$ 0.02 & 7.90 $\pm$ 0.02 \\

\enddata
\end{deluxetable}

Uncertainties in the flux densities are taken to be $\sigma\sqrt{N_b}$
where $\sigma$ is the rms noise level in the image and~$N_b$ is the
size of the region expressed as a number of synthesized beams.  These
uncertainties do not take into account various systematic effects,
e.g., the amount of flux resolved out by the interferometer, so should 
be viewed conservatively as lower limits to the actual uncertainties.

A division into only eastern and western halves of the remnant is
clearly a crude measure of the morphology.  Nonetheless, it will serve
our purpose of demonstrating a link between the morphology and
spectrum, and it will suffice for obtaining an estimate of the amount
of absorption along the line of sight (see below).

Following \cite{k89} we fit the spectra for the eastern and western
halves of the remnant to the functional form
\begin{equation}
S_\nu
 = S_{408}\left(\frac{\nu}{408\,\mathrm{MHz}}\right)^\alpha e^{-\tau_{408}(\nu/408\,\mathrm{MHz})^{-2.1}}
\label{eqn:flux}
\end{equation}
where $S_{408}$ and $\tau_{408}$ are the flux density and free-free optical
depth, respectively, at the fiducial frequency of~408~MHz.

The best-fit values for the free-free optical depths, scaled to 74~MHz
assuming a $\nu^{-2.1}$ dependence, are $\tau_{74,\mathrm{east}} =
0.7$ for the eastern half and $\tau_{74,\mathrm{west}} = 1.6$ for the
western half.  The western half of \objectname[]{W49B} shows a
significant increase in optical depth as compared to the eastern half.

The value for~$\tau_{74,\mathrm{east}}$ is likely to be an
overestimate. As \cite{k89} discusses, if the low-frequency turnover
occurs near the lowest observation frequency, the resulting estimates
of the optical depth will be highly uncertain. He illustrates this
point by comparing Culgoora and Clark Lake Radio Observatory flux
density measurements for~\objectname[]{W49B}. However, given that the
flux density for the western half of the remnant at~74~MHz is
significantly below the value at~327~MHz, we regard
$\tau_{74,\mathrm{west}} = 1.6$ as being a robust estimate of the
free-free optical depth.  The equivalent emission measure for the
western half is
$\mathrm{EM} =
10^{4.3}\,\mathrm{cm}^{-6}\,\mathrm{pc}\,(T_e/10^4\,\mathrm{K})^{1.35}$,
for an electron temperature~$T_e$ of~$10^4$~K and where we have taken
the Gaunt factor to be unity.

\subsection{Intrinsic \textit{vs.} Extrinsic Absorption}\label{sec:extrinsic}

Intrinsic thermal absorption at low radio frequencies has been
detected toward two young Galactic SNRs, but we do not favor this
scenario as an explanation for the low-frequency absorption toward
\objectname[]{W49B}.  In \objectname[]{Cas~A} \citep{kpde95} the
morphology of the absorption is centrally condensed, consistent with
the presumed location of unshocked ejecta interior to the reverse
shock. Moreover, the magnitude of the absorption ($\tau_{74} < 0.1$)
is much less than what we observe toward the western side of
\objectname[]{W49B}. The observed absorption in the \objectname[]{Crab
Nebula} \citep{bkfpeh97} is not centrally condensed but has been
linked to one of the largest thermal filaments at the foreground
periphery of the remnant.  Furthermore, the degree of absorption
corresponds to an optical depth $\tau_{74} \ll 1$, even lower than
that seen within \objectname[]{Cas~A}.

\cite{fujimotoetal95} find a centrally-condensed X-ray morphology for
\objectname[]{W49B}, which they attribute to thermal emission from hot
gas in the remnant's interior.  This hot gas is unlikely to contribute
to any appreciable free-free absorption at our observing frequencies.
Using the parameters they derive for the hot gas ($T_e \sim
10^{7.4}$~K, $n \sim 10$~cm${}^{-3}$, $l \sim 3$~pc), we find an
optical depth of $\tau \sim 10^{-6}$ at~74~MHz.

We consider both the asymmetry and magnitude of the observed
absorption toward \objectname[]{W49B} to be evidence in favor of extrinsic thermal absorption, in agreement with conclusions drawn from earlier low frequency measurements. Since \objectname[]{W49B} is a young SNR, we cannot rule out that unshocked ejecta near the center of the remnant may be contributing to the absorption, but it is unlikely to dominate.

\subsection{Morphology of the Absorption: Scale Sizes}\label{sec:size}

The gross appearance of the 74~MHz image (Figure~\ref{fig:74}) and the
analysis of \S\ref{sec:spectrum} suggests that the absorber toward
\objectname[]{W49B} is a discrete structure covering mainly the
western half of the source with a scale size of a few arcminutes.  We
now use Figures~\ref{fig:74} and~\ref{fig:327} to assess the existence
of smaller scale absorption structures as well.

\begin{figure}[tbh]
\begin{center}
 \mbox{\psfig{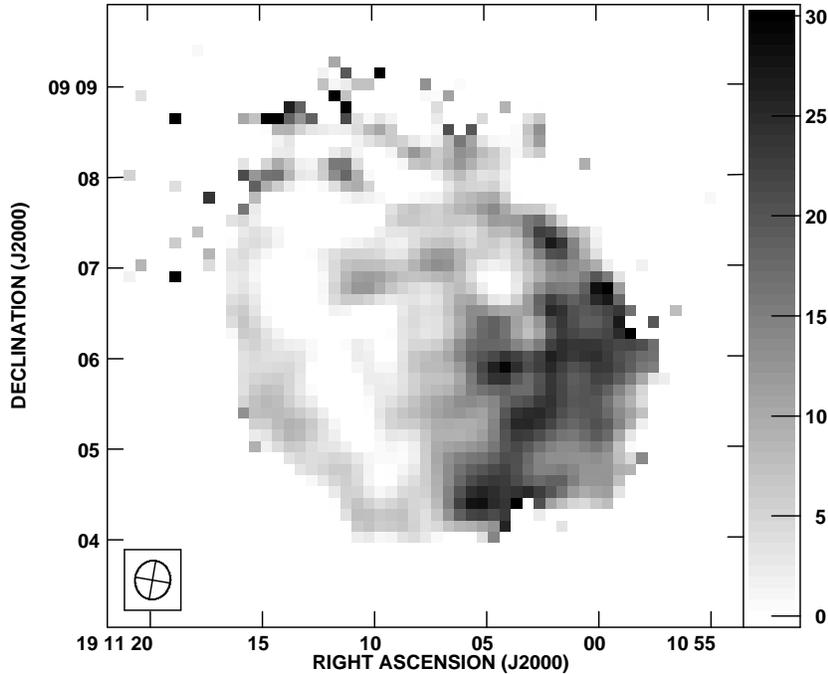}}
\end{center}
\vspace{-1.5cm}
\caption[]{The distribution of emission measure~EM across
the face of \protect\objectname[W]{W49B}.  This figure was constructed
by scaling the 327~MHz image (Figure~\ref{fig:327}) to~74~MHz and
forming an optical depth image based on the expected and observed
74~MHz images.  An electron temperature~$T_e = 10^4$~K and a Gaunt
factor of unity have been assumed.  The gray scale is linear between
0~pc~cm${}^{-6}$ and~$30 \times 10^3$~pc~cm${}^{-6}$.  Saturated
values along the rim of the remnant, particularly to the northeast,
result from regions of low signal-to-noise ratio.  The 74~MHz beam is
shown in the lower left.}
\label{fig:em}
\end{figure}

Assuming that the high-frequency spectrum can be extrapolated
to~74~MHz and that there are no intrinsic spatial variations in the
radio spectrum of the remnant between~74 and~327~MHz, we used the
327~MHz image to produce an expected 74~MHz
image~$I_{\mathrm{exp}}(\alpha, \delta)$.  In combination with the
observed 74~MHz image~$I_{\mathrm{obs}}(\alpha, \delta)$, we then
produced an image of the 74~MHz optical depth, $\tau_{74}(\alpha,
\delta) = -\ln(I_{\mathrm{exp}}/I_{\mathrm{obs}})$, and
$\mathrm{EM}(\alpha, \delta)$ across the face of the remnant.  A
temperature of~$10^4$~K and a Gaunt factor of unity were used to
convert $\tau_{74}$ to \hbox{EM}.  Figure~\ref{fig:em} shows the
resulting distribution of EM across the face of \objectname[]{W49B}.

Although the signal-to-noise level is low in the region of maximum EM
and optical depth (typically 3--$5\sigma$), Figure~\ref{fig:em}
clearly shows that there are small scale variations in EM across the
face of the remnant.  The typical scale size of the variations appears
to be comparable to or smaller than our beam, indicating that the
absorber has EM variations at least as small as 25\arcsec\ (1.2~pc
at~10~kpc).

\section{An Extended Hydrogen Envelope Toward W49B}\label{sec:discussion}

Extended hydrogen envelopes (EHEs) were postulated by \cite{a85},
primarily to explain various RRL observations \citep[see
also][]{pd76,a85}.  The observations require EHEs to be low density
($\sim 1$--10~cm${}^{-3}$), extended ($\sim 50$--200~pc),
inhomogeneously-distributed \ion{H}{2} gas.  In this section we show
that the absorber in front of \objectname[]{W49B} is likely to be a
series of classical \ion{H}{2} regions around which an EHE has formed.

\subsection{Previous Observations and Constraints on EHEs Toward W49B}\label{sec:EHEs}

\cite{pd76} have already analyzed the line of sight toward
\objectname[]{W49B}, using both RRLs, which are unusual in the
direction of a Galactic SNR, and previous estimates of the
low-frequency optical depth toward \objectname[]{W49B}.  Assuming that
the RRLs and the free-free absorption arise from the same gas, they
constrained the ionized gas to have an electron density~$n_e \sim
3$~cm${}^{-3}$ and temperature~$T_e \sim 10^3$--$10^4$~\hbox{K}.  In
order to make these determinations, \cite{pd76} had to make an
assumption about the radial extent of the ionized region.  They argued
that the transverse extent of the region is roughly 15\arcmin,
equivalent to a linear extent of~45~pc at a distance of~10~kpc, and so
that plausible values for the radial extent were 10~pc--1~kpc, leading
to the values quoted.

\subsection{Low-Frequency Imaging of \ion{H}{2} Regions and Their Envelopes}\label{sec:thiswork}

Our observations show that the absorption across the western face of
\objectname[]{W49B} has a mean value of $\tau \approx 1.6$ and can be
explained by a thermal cloud of size of a few arcminutes (10~pc
at~10~kpc) with $\mathrm{EM} =
10^{4.3}\,\mathrm{cm}^{-6}\,\mathrm{pc}\,(T_e/10^4\,\mathrm{K})^{1.35}$.
Pankonin \& Downes'~(1976) analysis used a free-free optical depth
of $\tau_{74} = 0.52$, a value determined by \cite{ds75} from
spatially unresolved observations.  Our higher angular resolution
measurements show that the optical depth is larger than what
\cite{pd76} assumed, but only by a factor of a few.  For an observed
$\tau$ and assumed radial extent~$L$, $n_e$ and $T_e$ are constrained
to be $n_e/T_e^{0.675} \propto \sqrt{\tau/L}$.  Therefore, the somewhat
smaller value of $\tau$ employed by \cite{pd76} does not change their
conclusions substantively and can be accommodated by a slightly higher
value for~$n_e$ or slightly lower value for~$T_e$ or both.
Furthermore, the few arcminute scale size constrained from our
observations is within a factor of a few of their estimate of
15\arcmin\ for the size of the ionized region.

Anantharamaiah's~(1986) model for an EHE is low-density \ion{H}{2}
gas surrounding, either partially or fully, a classical \ion{H}{2}
region.  Figure~\ref{fig:em} shows that there are EM variations on
size scales of approximately the synthesized beam (25\arcsec).  We now
show that these regions of highest EM are likely to be classical
\ion{H}{2} regions around which an EHE (or overlapping EHEs) has
formed.

The regions of maximum EM in Figure~\ref{fig:em} have $\mathrm{EM}
\simeq 3 \times 10^4$~pc~cm${}^{-6}$ (for an assumed temperature~$T_e
= 10^4$~K) and size of roughly 25\arcsec.  (We refer only to the
regions of maximum EM in the \emph{interior} of the western half of
the \hbox{SNR}.  The rim of the SNR, particularly to the northeast,
shows artificially high values of EM due to low signal-to-noise
ratios.)  The linear extent of these regions is roughly 1~pc
(at~10~kpc), so the electron density within these regions is $n_e \sim
10^2$~cm${}^{-3}$.  The electron density within these regions also
cannot exceed $10^2$~cm${}^{-3}$ by too large of a factor.
Figure~\ref{fig:327} shows that \objectname[]{W49B}'s morphology
at~327~MHz is similar to that seen at higher frequencies.  Thus, these
EM variations must not arise from regions so dense as to cause
appreciable free-free absorption at~327~MHz.  A free-free optical
depth at~327~MHz of~$\tau_{327} < 0.1$ requires that an \ion{H}{2}
region have $\mathrm{EM} < 3 \times 10^4$~pc~cm${}^{-6}$, for an
assumed temperature~$T_e = 10^4$~\hbox{K}.  The pressure within such a
region would be $nT \gtrsim 10^{6}$~K~cm${}^{-3}$, well in excess of
the typical ISM pressure of $nT \gtrsim 3000$~K~cm${}^{-3}$
\citep{kh88}.  Such objects would have to be either extremely young or
immersed in a higher pressure environment.  All of these
properties---$\mathrm{EM} \gtrsim 10^4$~pc~cm${}^{-6}$, $n_e \sim
10^2$~cm${}^{-3}$, size of roughly 1~pc, and high pressure---are
consistent with these EM variations representing classical \ion{H}{2}
regions.

Figure~\ref{fig:em} shows that not only are there localized
enhancements in the EM, but that there is a diffuse, extended region
of lower EM surrounding these localized enhancements.  The diffuse,
extended regions of lower EM have a typical value
of~$10^4$~pc~cm${}^{-6}$.  We emphasize two aspects of the regions of
lower \hbox{EM}.  First, the regions of lower EM represent regions of
\emph{higher} signal-to-noise ratios than the localized enhancements.
The lower EM regions are more significant.  Second, the typical size
of the localized EM enhancements are comparable in size to the
synthesized beam while being separated by angular distances comparable
to or larger than the synthesized beam.  Thus, the regions of lower EM
represent areas of real, lower absorption as opposed to a resolution
effect.  If classical, density-bounded \ion{H}{2} regions were in front of
\objectname[]{W49B} but located so close together on the sky that our
resolution was just capable of separating them, regions of apparently
lower EM would exist simply because of spatial averaging by the
synthesized beam.  Our synthesized beam is small enough and the
localized EM enhancements are well separated enough that the regions
of lower EM represent actual absorption.

Figure~\ref{fig:em} is constructed assuming $T_e = 10^4$~\hbox{K}.
Pankonin \& Downes'~(1976) analysis indicates that EHEs likely have a
lower temperature, probably closer to~$10^3$~\hbox{K}.  A lower
assumed temperature implies that the EM in the extended, diffuse EM
regions is a factor of order $10^{1.35}$ lower or $\mathrm{EM} \sim
10^3$~pc~cm${}^{-6}$.  For a scale size of order 10~pc, the density
within these diffuse EM regions is then $n_e \lesssim
10$~cm${}^{-3}$.  
Thus, the extended, diffuse EM portion of
Figure~\ref{fig:em} represents the EHE of \cite{pd76} and \cite{a86}
while the localized EM enhancements are the classical \ion{H}{2}
regions.  The lower density and pressure of the EHE suggests that the
classical \ion{H}{2} regions are only partially radiation-bounded or that
they are expanding into a lower density region \citep[a.k.a.~the ``blister''
model, ][]{i78}.

\subsection{Associated Neutral and Molecular Material}\label{sec:comparison}

The RRLs observed toward \objectname[]{W49B} have LSR velocities
near~60~\kms.  Both \ion{H}{1} \citep{bt00} and a variety of molecular
species, including OH \citep{ckt87} and H${}_2$CO \citep{bwd82,wrw88},
also are seen in absorption at the same velocity toward the
\objectname[]{W49} complex.  Most of these observations have targeted
\objectname[]{W49A}, though some (notably those cited above) have
included \objectname[]{W49B}.  We use those observations that have
resolved \objectname[]{W49B} to show that the ionized material has
associated neutral and molecular material.  We emphasize that we shall
focus only on material at a velocity of~60~\kms; the \ion{H}{1}
absorption in particular is both spatially complex and present at a
number of velocities \citep{bt00}, but RRLs are seen only at~60~\kms.

\begin{figure}[tbh]
\begin{center}
 \mbox{\psfig{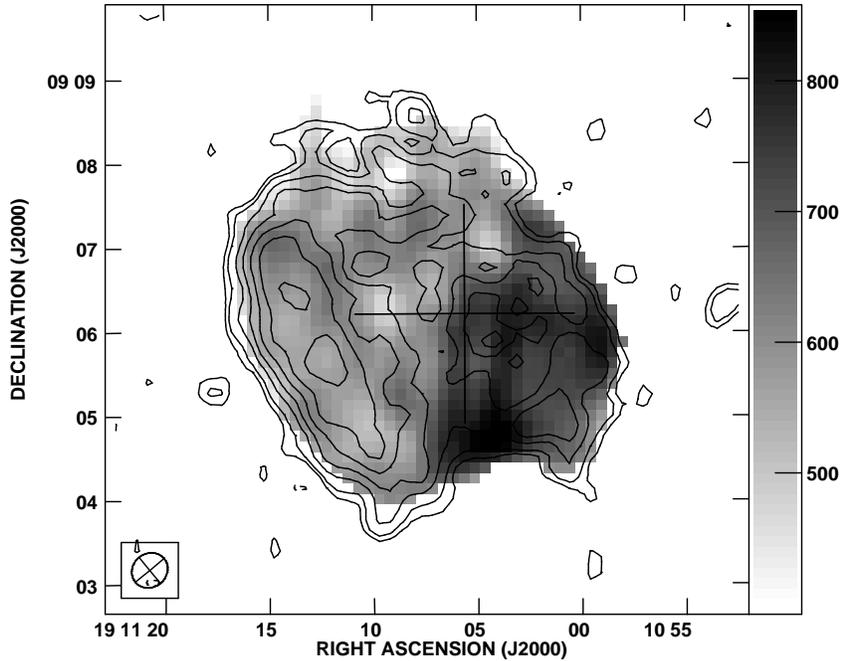}}
\end{center}
\vspace{-1.5cm}
\caption[]{A comparison of the 74~MHz radio continuum,
the \ion{H}{1} absorption at~60~\kms, and the H${}_2$CO absorption
at~60~\kms.  The 74~MHz continuum is shown in contours (as for
Figure~\ref{fig:74}) while the gray scale shows the \ion{H}{1}
absorption.  The gray scale levels are linear and range from a column
density of~$N_H/T_s = 4 \times 10^{19}$~\cd~K${}^{-1}$ to~$8.5 \times
10^{19}$~\cd~K${}^{-1}$.  The cross marks the location of maximum
H${}_2$CO absorption, and its size is approximately the size of the
beam (2\farcm6) used in those observations.  The \ion{H}{1} absorption
data are from \cite{bt00}, and the H${}_2$CO absorption was measured
by \cite{bwd82}.}
\label{fig:hi}
\end{figure}

Figure~\ref{fig:hi} shows the angular distribution of the \ion{H}{1}
absorption at~60~\kms\ measured by \cite{bt00} compared to the 74~MHz
radio continuum (Figure~\ref{fig:74}).  The close correspondence
between the region of maximum \ion{H}{1} absorption and the 74~MHz
continuum absorption is clear.  Moreover, \cite{bt00} describe the
appearance of the \ion{H}{1} absorption as ``spotty,'' which is
similar to the appearance of the 74~MHz image (Figure~\ref{fig:74}).
Observations with the Chandra X-ray telescope also suggest that the
X-ray emission is attenuated in the southwestern portion of the
remnant (Hwang~2000, private communication), consistent with an
increase in \ion{H}{1} column density along this line of sight.
Figure~\ref{fig:hi} also shows the location of the peak H${}_2$CO
absorption at~60~\kms\ measured by \cite{bwd82}.  Although Bieging et
al.'s~(1982) observations only marginally resolved \objectname[]{W49B}
(2\farcm6 beam), the peak H${}_2$CO absorption occurs on the western
side of the \hbox{SNR}.

The correspondence between the angular distribution of the \ion{H}{1}
and H${}_2$CO absorption at~60~\kms\ and the 74~MHz continuum
absorption and the common velocities of the RRLs, \ion{H}{1}
absorption, and H${}_2$CO absorption (and other molecular species)
indicates that the ionized, neutral, and molecular material is
spatially collocational along this line of sight.  Furthermore, the
appearance of both the ionized and neutral absorption suggests that
the material has considerable small-scale structure.  This material
probably is located within the Sagittarius arm \citep{gw94}.  If so,
it could be considerably closer (factor of~2 or more closer or~5~kpc) than
the presumed distance to \objectname[]{W49B}.  Though we have assumed
a distance of~10~kpc in converting angles to distances above, using a
distance of~5~kpc will lead to no more than a factor of~2 error in the
derived quantities (e.g., $n_e$).

Absorption in various molecular species is seen toward
\objectname[]{W49A} at a velocity of~60~\kms.  We consider it likely
that common material is responsible for the absorption toward both
\objectname[]{W49A} and \objectname[]{W49B} at~60~\kms.  Comparison of
line ratios and radiative transfer modeling suggest that the
molecular material is at a low density ($< 10^4$~cm${}^{-3}$) and has
an excitation temperature close to~3~\hbox{K}.  Many authors
\citep[e.g.,][]{wrw88,gw94} have concluded that these molecular
clouds do not currently support any star formation.

This apparent discrepancy between the 74~MHz observations---clearly
indicating the presence of ionized material---and the molecular
observations---suggesting no ionizing sources---may have a number of
resolutions.  The molecular clouds may simply have a low level of star
formation or the \ion{H}{2} regions may be produced by passing
early-type stars that are otherwise not associated with the clouds.

Cosmic-ray ionization is unlikely to be the explanation of this
discrepancy as the ionization rate required to produce the observed
electron density if the neutral and ionized material are mixed
\citep{htc71} is $10^{-13}$--$10^{-12}$~s${}^{-1}$, orders of
magnitude larger than the nominal cosmic ray ionization rate
\citep{kh88}.  Future observations of molecular species with a
resolution similar to or better than that of the 74~MHz and \ion{H}{1}
observations would aid greatly in assessing to what extent the
molecular material shows a similar morphology.

\subsection{Interstellar Scattering toward W49B}\label{sec:scattering}

Lines of sight to the west of \objectname[]{W49B} show enhanced
radio-wave scattering, via enhanced angular broadening measurements of
OH masers and extragalactic sources \citep{km82,fsc91}.  The gas
responsible for the RRLs and free-free absorption may
be responsible for this enhanced scattering.

It is commonly assumed that the density fluctuations responsible for
interstellar scattering have a power-law power spectrum parameterized as
\begin{equation}
P_{\delta n_e}(q, z) = C_n^2(z)q^{-\alpha} 
\end{equation}
for length scales~$q^{-1}$ between an inner scale~$l_1$ and an outer
scale~$l_0$ \citep{r90}.  The (slowly-varying) coefficient~$C_n^2$
describes the strength of density fluctuations and is related to the rms
density.  Analogous to the emission measure~EM, the scattering
measure~SM is
the line-of-sight integral of $C_n^2$, 
\begin{equation}
\mathrm{SM} = \int dz\,C_n^2(z).
\end{equation}
Various observables \cite[for a review see][]{tc93}, including
scattering diameters, can be used to infer \hbox{SM}.  Based on the EM
of the EHE, the scattering measure contributed by this absorbing cloud
can be no more than $\mathrm{SM}
\lesssim 7\,\mathrm{kpc}\,\mathrm{m}^{-20/3}\,(l_0/10\,\mathrm{pc})^{2/3}$
\citep{cwfsr91}, for an assumed outer scale of $l_0 = 10$~pc.  The SM
must be less than this value because all free electrons along the line
of sight contribute to the EM while only fluctuations in the electron
density contribute to \hbox{SM}.

The measured angular diameters imply $\mathrm{SM} \sim
3$~kpc~m${}^{-20/3}$.  Taken at face value our predicted value for the
SM is sufficient to explain the enhanced scattering.  However, we have
assumed a larger than normal value for~$l_0$.  In particular, other
analyses of enhanced scattering regions suggest that $l_0 \ll 1$~pc
\citep{fkv93,lc98}.

In general, unambiguous association of an absorbing cloud with the
scattering material will be difficult.  \objectname[]{W49B} (and many
other supernova remnants) lies at a fairly large distance at a low
Galactic latitude toward the inner Galaxy.  Other ionized gas along
the line of sight could contribute to the radio-wave scattering.

\section{Conclusions}\label{sec:conclude}

We have used the VLA to image the Galactic supernova remnant
\objectname[]{W49B} at~74 and~327~MHz with~25\arcsec\ and~6\arcsec\
angular resolution, respectively.  Our 74~MHz image
(Figure~\ref{fig:74}) marks the lowest frequency at which this remnant
has been angularly resolved.  At frequencies above 1~GHz,
\objectname[]{W49B} has a limb-brightened shell.  This remnant retains
this morphology at~327~MHz (Figure~\ref{fig:327}).

A dramatic difference in the remnant's morphology is seen
between~74~MHz and higher frequency images, with the southwestern
portion of the remnant being significantly fainter.  The flux
densities we measure at~74 and~327~MHz are consistent with previous
determinations of a low-frequency turnover in the SNR's integrated
continuum spectrum near~80~MHz.  We have also compared the spectra of
the eastern and western halves of the SNR (Figure~\ref{fig:spectrum}).
The western half of the SNR is significantly more absorbed,
demonstrating that the morphological differences and the low-frequency
turnover are linked.

Both the low-frequency turnover and morphology of \objectname[]{W49B}
are due to foreground ionized gas.  Earlier analyses \citep{pd76,a86},
based on lower resolution observations, suggested that the ionized gas
should be extended ($\sim 10$~pc or a few arcminutes) and low density
($\sim 3$~cm${}^{-3}$).  The gross morphology and integrated spectrum
that we obtain are consistent with these analyses, if we include
additional absorption from small scale \ion{H}{2} regions as well.

Our 74 and~327~MHz images have sufficient angular resolution to enable
us to resolve the distribution of emission measure~EM across the face
of \objectname[]{W49B} (Figure~\ref{fig:em}).  The EM distribution
shows localized enhancements---$\mathrm{EM} \gtrsim
10^4(T_e/10^4\,\mathrm{K})^{1.35}$~pc~cm${}^{-6}$, $n_e
\sim 10^2$~cm${}^{-3}$, size of roughly 1~pc, and high pressure---that
we identify as being classical \ion{H}{2} regions.  Strengthening
their identification as \ion{H}{2} regions is the high correlation
between the distribution of EM and \ion{H}{1} absorption.  Regions of
high EM are also regions of high \ion{H}{1} absorption.  At the same
velocity as the \ion{H}{1} absorption, there are also H${}_2$CO
absorption observations that just resolve the \hbox{SNR}.  These
molecular observations also indicate an increase in absorption to the
western side of the \hbox{SNR}.

Surrounding the localized EM enhancements are regions of lower
\hbox{EM}.  This lower EM material represents the extended,
low-density, ionized gas or extended \ion{H}{2} envelope suggested by
previous analyses.  We have therefore resolved the low-frequency
absorption in front of \objectname[]{W49B} into a complex of
\ion{H}{2} regions and an EHE (or nearly overlapping EHEs).

The ionized gas responsible for the free-free absorption and radio
recombination lines may be responsible for the enhanced radio-wave
scattering seen in the general direction of the \objectname[]{W49}
complex.  However, given the line of sight into the inner Galaxy, it
is difficult to make an unambiguous determination of the location of
the enhanced scattering region.

Future observations at both low and high frequencies will be essential
for constraining further the properties of the ionized gas responsible
for low-frequency turnovers.  Future high-frequency interferometers
may have sufficient sensitivity to map the angular distribution of
molecular material toward \objectname[]{W49B} and other supernova
remnants.  At low frequencies resolving the angular distribution of
the radio recombination lines is essential to verifying a crucial
assumption in both this work and that of \cite{pd76}, that the RRLs
and the free-free absorption do arise from the same ionized gas.
Low-frequency observations of other supernova remnants are planned
with both the NRL-NRAO 74~MHz system at the VLA (and the GMRT and
other future low-frequency interferometers) and will be used to
establish to what extent the line of sight toward \objectname[]{W49B}
is or is not representative.  Low-frequency observations of a
high-latitude or outer Galaxy supernova remnant may also allow a more
definitive connection to be established between the extended
low-density ionized gas and the gas responsible for radio-wave
scattering.  Particularly powerful would be a demonstration of a
correlation in free-free absorption and scattering, e.g., enhanced
angular broadening for sources seen near one side of a remnant showing
enhanced free-free absorption.

\acknowledgements

We thank W.~Erickson for illuminating discussions, D.~Moffett and
S.~Reynolds for making their higher frequency images available, and
C.~Brogan for making her \ion{H}{1} data available. This research made
use of NASA's ADS Abstract Service and the SIMBAD database, operated
at the CDS, Strasbourg, France. The National Radio Astronomy
Observatory is a facility of the National Science Foundation operated
under cooperative agreement by Associated Universities, Inc. This work
was conducted while CKL held a National Research Council-NRL Research
Associateship. Basic research in radio astronomy at the NRL is
supported by the Office of Naval Research.

\end{document}